\documentclass[conference]{IEEEtran}
\usepackage{cite}
\usepackage{amsmath,amssymb,amsfonts}
\usepackage{algorithmic}
\usepackage{graphicx}
\usepackage{subfigure}
\usepackage{textcomp}
\usepackage{xcolor}
\def\BibTeX{{\rm B\kern-.05em{\sc i\kern-.025em b}\kern-.08em
		T\kern-.1667em\lower.7ex\hbox{E}\kern-.125emX}}

\usepackage{amsthm}
\usepackage[ruled]{algorithm2e}
\usepackage{mathrsfs}
\usepackage{bm,epsfig,amsthm,url}
\usepackage{indentfirst}
\usepackage{epstopdf}
\usepackage[top=0.75in, left=0.7in, bottom=1.1in, right=0.7in]{geometry}

\renewcommand{\raggedright}{\leftskip=0pt \rightskip=0pt plus 0cm}

\theoremstyle{definition}

\begin{document}
	
	\title{UAV-Assisted Over-the-Air Computation}

\author{
	\IEEEauthorblockN{$\text{Min Fu}^{\ast}$, $\text{Yong Zhou}^{\ast}$,  $\text{Yuanming Shi}^{\ast}$, $\text{Ting Wang}^{\dagger}$, and $\text{Wei Chen}^{\S}$
	}\\
	\IEEEauthorblockA{
		$^{\ast}$School of Information Science and Technology, ShanghaiTech University, Shanghai 201210, China \\ 
	$^{\dagger}$Shanghai Key Lab. of Trustworthy Computing, School of Software Engineering,	East China Normal University\\
	$^{\S}$Department of Electronic Engineering, Tsinghua University, Beijing 100084, China\\
		Email:  \{fumin, zhouyong, shiym\}@shanghaitech.edu.cn, twang@sei.ecnu.edu.cn, wchen@tsinghua.edu.cn}
}
	\maketitle
	\setlength\abovedisplayskip{2pt}
	\setlength\belowdisplayskip{2pt}
	\vspace{-1.2cm}
	\begin{abstract}
Over-the-air computation (AirComp) provides a promising way to support ultrafast aggregation of distributed  data. However, its performance cannot be guaranteed in long-distance transmission due to the distortion induced by the channel fading and noise.
	To unleash the full potential of AirComp, this paper proposes to use a low-cost unmanned aerial vehicle (UAV) acting as a mobile base station to assist AirComp systems. 
	Specifically, due to its controllable high-mobility and high-altitude, the UAV can move sufficiently close to the sensors to enable line-of-sight transmission and  adaptively adjust all the  links' distances, thereby enhancing the signal magnitude alignment and  noise suppression.
	Our goal is to minimize the time-averaging mean-square error for AirComp  by jointly optimizing the UAV trajectory, the scaling factor  at the UAV, and the transmit power at the sensors, under constraints on the UAV's predetermined locations and  flying speed, sensors' average and peak power limits.	
	However, due to the highly coupled optimization variables and time-dependent constraints, the resulting problem is non-convex  and challenging.
	We thus propose an efficient iterative algorithm by applying the block coordinate descent and successive convex optimization techniques.
	Simulation results verify the convergence of the proposed algorithm and demonstrate the  performance gains and robustness of the proposed design  compared with  benchmarks.
	\end{abstract}
\section{Introduction}\label{introduction}
In the future Internet-of-Things (IoT) based big data applications, both the ultrafast data collection from massive sensors with limited spectrum bandwidth and effective interpretation on collected data  with limited computation capacity are highly challenging.
For example, in the scenario of environmental monitoring, multiple sensors distributed over a particular area concurrently transmit their measured environmental data (e.g., temperature and humidity), while the monitor needs to receive and compute the average value of  the measured data.
To tackle these issues, over-the-air computation (AirComp)  has recently been proposed as a promising multiple access scheme, which integrates the computation into communication  \cite{Zhu2020c}, \cite{Yang2020FL}. 
The basic principle of AirComp is to exploit the waveform superposition property of multiple-access channel (MAC) to compute a class of nomographic functions (e.g.,  mean and weighted sum)  of numerous data  via concurrent transmission.
Furthermore, 
this can enable a series of  IoT applications ranging from latency-sensitive  sensing \cite{Jiang2019, Li2019, Dong2020BlindAirComp} to data-intensive federated machine learning \cite{Yang2020FL},\cite{Letaief2019a}.

In practice, the  performance of AirComp is significantly limited by the signal distortion due to channel fading and noise, especially in long-distance transmission. 
Note that the noise power is comparable to the signal power in long-distance transmission due to analog transmission in most AirComp systems. 
To enable reliable AirComp, the key designs in AirComp are the  power control at sensors for coping with channel attenuation and signal scaling  at the base station (BS) for noise suppression \cite{Li2019,Cao2020Optimized}.
In prior works, uniform channel inversion is  implemented at the sensor to perfectly align  the magnitude of signals \cite{Yang2020FL, Jiang2019, Li2019}.
However, this scheme may severely degrade the AirComp performance when one or more individual channels are in deep fading, which will amplify the negative effect of noise.
Recently, the authors in \cite{Cao2020Optimized}  proposed an optimal policy with threshold-based structure to combine channel inversion and full power transmission.
As shown in \cite{Cao2020Optimized}, enlarging noise suppression can lead to an increased signal misalignment error  since the sensors are usually   power constrained.
Therefore, due to the channel fading and noise as well as limited transmit power at sensors, 
 relying only on the terrestrial BSs may not be able  to guarantee the performance of AirComp in long-distance transmission, especially when the BSs are sparsely deployed or unavailable (e.g., wild-area monitoring applications).

As a remedy to the above limitations, low-cost unmanned aerial vehicle (UAV)  is considered as a promising alternative to assist the terrestrial networks \cite{Wang2018b,Wang2019a, Zeng2019}.
Recently, the research efforts  have been devoted to employing UAVs as  mobile BSs in  IoT networks, such as  information dissemination \cite{Zeng2017EnergyEfficient, Wu2018MultiUAV}, one-by-one data collection \cite{Zhan2019}. 
In this paper, we shall propose to deploy the UAV-mounted BS to assist AirComp, thereby enjoying the following advantages compared with the conventional  BS. 
First, the UAV-mounted BS is cost-effective and can  move sufficiently close to the sensors even in the wild area, which can avoid long-distance transmission and thus save the sensors' power and mitigate the effect of noise.
Second, due to its high altitude, UAV usually have line-of-sight (LoS) connection with ground sensors, which can reduce the probability of channels in deep fading and thus enhance signal magnitude alignment.
Furthermore, the controllable high-mobility  UAV can  actively construct favorable channels to  strike a balance between communication distance and sensors'  heterogeneous power constraints.
Specifically, UAV can dynamically adapt its trajectory to fly closer to sensors with lower power than these with higher power to align the magnitude of signals, which possess an additional degree of freedom for the AirComp performance enhancement.
Hence, this motivates us to study a new AirComp technique referred to as UAV-assisted AirComp, which has the potential to overcome  the   issues of conventional AirComp.

In this paper, we consider a UAV-assisted AirComp system, in which  the UAV acting as a mobile BS is dispatched to collect average information of distributed data generated by ground sensors via AirComp during a given mission interval. 
A common design metric that has been widely adopted in AirComp is the  mean-squared error (MSE) between the estimated function value and the target function value \cite{Yang2020FL, Jiang2019, Cao2020Optimized,  Li2019}.
Hence,  our goal is to minimize the time-average MSE by jointly optimizing the UAV trajectory, signal scaling factor (termed as denoising factor) at UAV, and transmit power at ground sensors.
However, due to the highly coupled variables and time-dependent constraints, it is challenging to solve the resulting problem optimally in general.
To address these  challenges, we develop an efficient iterative algorithm by applying the block coordinate descent  method, where the transmit power, denoising factor, and UAV trajectory are optimized in an alternating manner.
However, the UAV trajectory optimization subproblem is still difficult to be solved due to the non-convexity of its objective function, for which the successive convex approximation  is applied to solve it approximately. 
The numerical results validate the convergence of the proposed algorithm.
It is also observed that  UAV trajectory design can dynamically adapt to different transmit power, the proposed joint design can achieve significant performance gains and the robustness, as compared to the benchmarks.

\section{System  Model and Problem Formulation}\label{model}
\begin{figure}[t]
	\centering
	\includegraphics[scale=0.3]{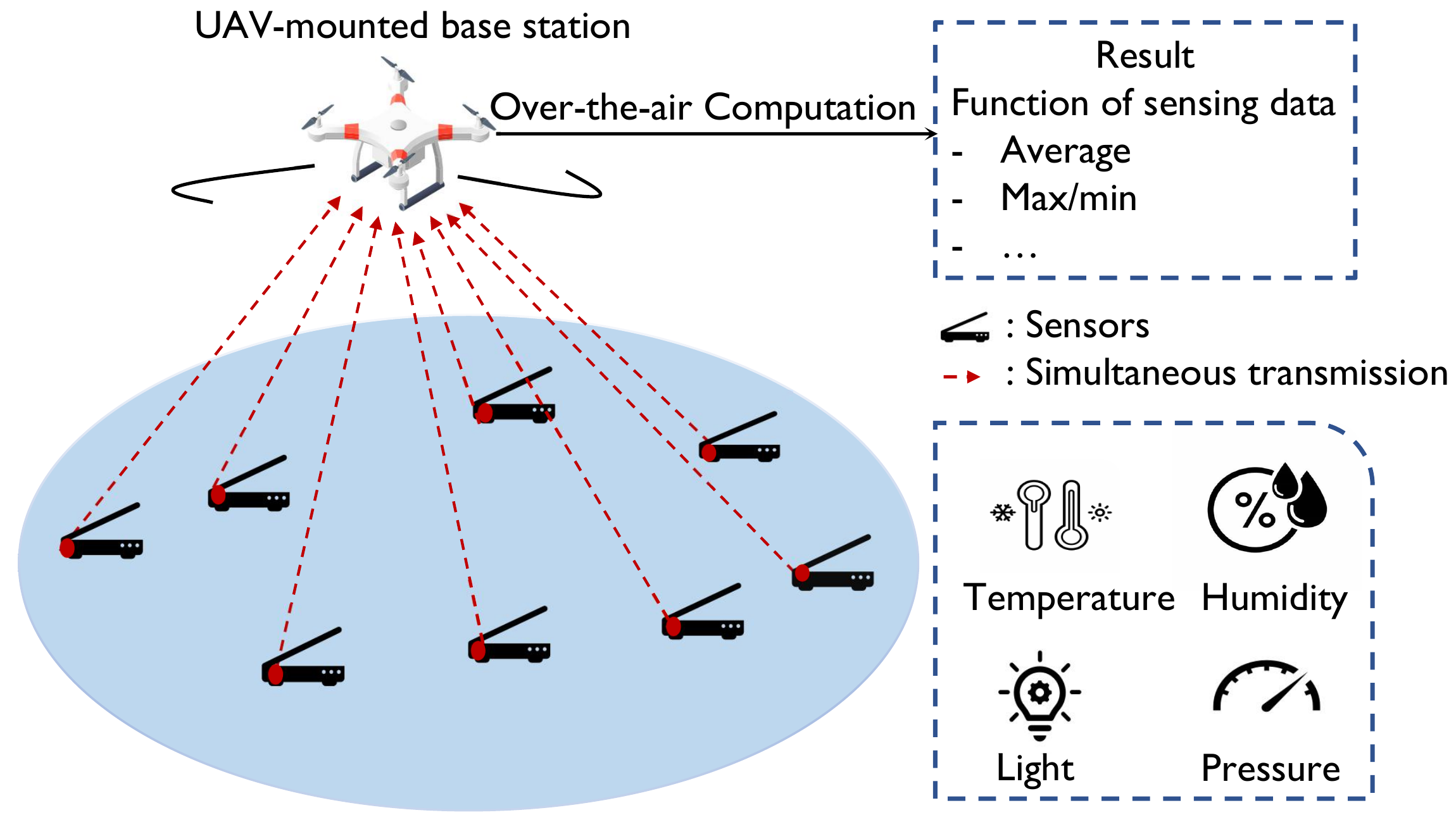}
	\caption{UAV-assisted AirComp networks}\label{F:Systemmodel}
	\vspace{-5mm}
\end{figure}
As illustrated in Fig. \ref{F:Systemmodel},  we consider a UAV-assisted AirComp network with $K$ ground sensors.
Both the UAV and  sensors are equipped with a single antenna due to their size and power limitations.
The mobile UAV  is interesting in exploiting the AirComp to aggregate the average function of distributed data generated by $K$ ground sensors during a given flight time  $T$ second (s) rather than obtaining the value of each individual data.
Therein, due to its  controllable high-mobility and high-altitude, UAV can move sufficiently close to sensors with LoS connections and  adaptively balance between communication distance and sensors' transmit power, which can save sensors'  power  for the signal magnitude alignment and suppress noise.
One practical scenario for such a consideration could be wild-area environmental monitoring when the BSs are unavailable nearby, where the UAV is employed to  monitor the average value of the temperature  measured by  sensors distributed over a particular area. 
Note that we focus on the basic scenario with a single UAV and without conventional terrestrial BS. The extension to consider more general cases with multiple cooperative UAVs and/or in the presence of ground BSs will be left as our future work.

We consider a three-dimensions (3D) Cartesian coordinate system, where the horizontal coordinate of sensor $k$ is denoted as  $\mathbf w_k=[x_k, y_k]\in \mathbb{R}^{1 \times 2}$ with $x_k$ and $y_k$ being  $x$- and $y$- coordinates, respectively. The set of ground sensors is denoted as  $\mathcal {K}\triangleq \{1,\ldots,K\} $, $K> 1$.
During the flight of the UAV, the ground sensors' locations are assumed to be fixed and known to the UAV.
In practice, the locations of  sensors could be determined by the standard positioning techniques (e.g.,  GPS based localization).
Additionally, we assume that the UAV flies at a fixed altitude $H$ above the ground level. Note that in practice, $H$ corresponds to the minimum altitude that ensures obstacle avoidance without the need for frequent aircraft ascending and descending.

\subsection{UAV Trajectory Model and Channel Model}
We denote the time-varying UAV trajectory projected on the horizontal plane as $\bm{q}(t)=[x(t), y(t)]\in \mathbb{R}^{1\times 2}$, $0\leq t\leq T$.
We assume that UAV starts the mission at a predetermined location, the coordinate is denoted as $[x_0, y_0, H]\in \mathbb{R}^{1\times 3}$ while it needs to return to the same location after the mission completes. 
Note that in practice, the predetermined location is determined according to various aspects, e.g., replenishing energy and/or offloading the computation data \cite{Zhan2019}, \cite{Group2020}.
 We denote $\bm q(0) = [x_0, y_0]$ and $\bm q(T) = [x_0, y_0]$. 
 Besides, we denote the  maximum speed of the UAV as $V_{\rm{max}}$ in meter/second (m/s).
 Hence, we have the constraints
$\sqrt{\dot{x}^2(t) +\dot{y}^2(t)} \leq V_{\text{max}}$, $0 < t < T$, where
$\dot{x}(t)$ and $\dot{y}(t)$ denote the time-derivatives of $x(t)$ and $y(t)$ at time instant $t$, respectively.

To assist a tractable algorithm design, we adopt the time discretization technique to deal with the continuous UAV trajectory design, which is widely considered in most of the existing works\cite{Wu2018MultiUAV, Zeng2017EnergyEfficient,Zhan2019,Group2020}.
 Specifically, the mission duration $T$ is equally divided into $N$ time slots, i.e., $ T = N\delta $, where $\delta$ denotes the time step size.
Given the maximum UAV speed $V_{\max}$ and altitude $H$,  the time step size $\delta$ needs to be carefully chosen so that the distance between the UAV and the  sensors is approximately constant during each time slot, i.e., $\delta V_{\text{max}}\ll H$.
Therefore, the UAV trajectory $\bm q(t)$ over time horizon $T$ is approximated by the $N$-length sequence $\{\bm{q}[n]\}_{n =1}^{N}$ with $\bm{q}[n] \triangleq \bm q(n\delta) $ denoting the UAV's horizontal coordinate at time slot $n$.  
 To this end,  the UAV's mobility constraints can be written as
 \setlength\arraycolsep{1.2pt}
\begin{eqnarray}
\|\bm q [n]-\bm q[n-1]\|_2 &\leq& V_{\max}\delta,  n = 1, \ldots, N,  \label{speed constr}\\
\bm{q}[0] &=& [x_0, y_0],  \label{initial location}\\
\bm{q}[N] &=& [x_0, y_0],  \label{final location}
\end{eqnarray} 
where constraints \eqref{speed constr} correspond to the UAV speed constraint and constraints \eqref{initial location} and \eqref{final location} are subject to the predetermined locations, respectively.
 Note that a smaller value of $\delta$ makes the discrete-time approximation more accurate while inevitably increasing the complexity of the trajectory design. 
 Thus, the time step size $\delta$ needs to be  properly chosen to strike a balance between design accuracy and design complexity.

Recent field experiments by Qualcomm have verified that the UAV-to-ground channel is indeed dominated by the LoS link for UAV flying above a certain altitude \cite{Qual2017LET}. 
Therefore, we assume that each communication link from  the  sensor to the UAV is dominated by the  LoS channel.
Moreover, the Doppler effect resulting from the UAV mobility is assumed to be perfectly compensated \cite{Mengali1997Sync}. 
Thus, the time-varying channel from the  sensor $k$ to  the UAV at time slot $n$ is modeled as
\begin{eqnarray}
{h}_k[n]=\sqrt{\beta_k[n]}\tilde{h}_k[n],\label{chan0}
\end{eqnarray}
where $|\tilde{h}_k[n]| =1$  and $\beta_k[n]$ denotes the large-scale channel power gain due to the free-space path loss, as in \cite{Wu2018MultiUAV}. Specifically, $\beta_k[n]$ is modeled as
$
\beta_k[n]=\beta_0 d_k^{-\alpha}[n],\label{chan1}
$
where $\beta_0$ represents the channel power gain at the reference distance of $d_0=1$ m related to the carrier frequency and antenna gain,  $\alpha \geq 2$ is the path loss exponent, and $d_k[n] = \sqrt{H^2+\|\bm{q}[n]-\textbf{w}_k\|_2^2}$ is the  distance between UAV and sensor $k$ at time slot $n$.

\subsection{AirComp for Mobile Aggregation}
The UAV aims to compute a target function of the aggregated data from all ground sensors.
Let $Z_{k}[n] \in \mathbb{C}$ denote the data measured by sensor $k$ at  time slot $n\in{\cal N}$. The target function computed at the UAV can be written as
\begin{eqnarray}
f[n]=\phi \Big(\sum \limits_{k\in{\cal K}}\psi_k(Z_{k}[n])\Big),\label{functions}	
\end{eqnarray}
where $\phi$ is the post-processing function at UAV, $\psi_k$ is the pre-processing at sensor $k$.
Denote $s_k[n] \triangleq \psi_k(Z_{k}[n])$ as the  symbols transmitted at sensor $k$. The symbols are assumed to be independent and normalized with zero mean and unit variance, i.e., $\mathbb{E}(s_k[n]) = 0$, $\mathbb{E}(s_k[n]s_k^{\sf H}[n]) = 1$, and $\mathbb{E}[s_{i}[n]s_{j}[n]^{\sf H}] = 0 , \forall i \neq j$, as in \cite{Yang2020FL}, \cite{Cao2020Optimized}.

Without loss of generality, in this paper, we consider the case where the UAV  computes the average of distributed data generated by  sensors \cite{Yang2020FL}, \cite{Cao2020Optimized}.
Therefore, the  function of interest at the UAV at time slot $n$ is given by
\begin{eqnarray}
f[n]=\frac{1}{K}\sum \limits_{k\in{\cal K}}s_{k}[n],\label{desired_functions}
\end{eqnarray}
where $1/K$ is the post-processing function at UAV.
Our goal is to recover this target function $f[n]$ by exploiting the superposition property of  MAC, i.e., AirComp.

At time slot $n$, the received signal at the UAV is given by
\begin{eqnarray}
y[n]=\sum \limits_{k\in{\cal K}}b_k[n]h_{k}[n]s_{k}[n]+ e[n],\label{time_Y}
\end{eqnarray}
where $b_k[n] \in \mathbb{C}$ denotes the  transmit precoding coefficient at sensor $k$  and $e[n]$ denotes the  additive white Gaussian noise (AWGN), i.e., $e[n]\thicksim\cal{C}\cal{N}$$(0,\sigma^2)$.
The transmit power constraint at sensor $k$ is given by
\begin{eqnarray}
	\mathbb{E}(|b_k[n]s_k[n]|^2) = |b_k[n]|^2 \leq P_k,
\end{eqnarray}
where $P_k> 0$ is the maximum transmit power of sensor $k$. In addition, we consider that each sensor has  the following average transmit power constraint
\begin{eqnarray} 
	\frac{1}{N}\sum_{n=1}^{N} |b_k[n]|^2 \leq \bar{P}_k , \forall k.\label{average p}
\end{eqnarray}
To make  constraint  \eqref{average p} non-trivial, we set $\bar{P}_k \leq P_k$.

Upon receiving  signal $y[n]$ in \eqref{time_Y}, the estimated average of transmitted data at UAV is given by
\begin{eqnarray}\label{ave_functions}
\hat f[n] = \frac{y[n]}{K\sqrt{\eta[n]}},
\end{eqnarray}
where  $\eta[n]$ is a denoising factor for noise suppression \cite{Yang2020FL}, \cite{Cao2020Optimized}.

\subsection{Performance Metric}
We are interested in minimizing the MSE between the estimation function $\hat f[n]$ and the desired function $f[n]$, which is widely adopted to quantify the signal distortion in most of existing AirComp works \cite{Yang2020FL, Jiang2019, Cao2020Optimized,  Li2019}.
In particular, the corresponding  MSE at time slot $n$ is given by
\begin{eqnarray}
&&{\rm MSE}[n]=\mathbb{E}[|\hat{f}[n]-f[n]|^2]\nonumber \\
&&=\frac{1}{K^2}\mathbb{E}\Bigg[\Bigg(\frac{y[n]}{\sqrt{\eta[n]}}- \sum \limits_{k\in{\cal K}}s_{k}[n]\Bigg)^2 \Bigg] \nonumber \\
&& = \frac{1}{K^2}\Bigg(\sum \limits_{k\in{\cal K}}\Bigg(\frac{b_k[n]h_{k}[n]}{\sqrt{\eta[n]}}-1\Bigg)^2+\frac{\sigma^2}{\eta[n]}\Bigg),
\end{eqnarray}
where the expectation is taken over the distributions of the transmitted signals $\{s_k[n]\}$ and  noise.

For simplicity, we only focus on the  power control at each sensor and let $b_k = \frac{ \sqrt{p_{k}[n]} h_{k}^{\dagger}[n]}{ |h_{k}[n]|}$,
where $p_k[n]\!\ge \!0$ denotes the transmit power at sensor $k\in\! \cal K$ at  time slot $n$ and ${\dagger}$ denotes the conjugate operation. 

Hence, the MSE is given as
\begin{eqnarray}
\!\!\!\!\!&&{\rm MSE}[n]  =  \frac{1}{K^2}\Bigg(\sum \limits_{k\in{\cal K}}\Bigg(\frac{\sqrt{p_{k}[n]}\big|h_k[n]\big|}{\sqrt{\eta[n]}}-1\Bigg)^2+\frac{\sigma^2}{\eta[n]}\Bigg) \nonumber \\
\!\!\!\!\!&&\!=\! \frac{1}{K^2}\!\Bigg(\!\sum \limits_{k\in{\cal K}}\!\left(\!\frac{\sqrt{p_{k}[n]}\sqrt{\beta_0}}{\sqrt{\eta[n]}(H^2+\|\bm{q}[n]-\textbf{w}_k\|_2^2)^{\frac{\alpha}{4}}}\!-1\!\right)^2 \!+\!\frac{\sigma^2}{\eta[n]}\!\Bigg).\nonumber \\
\end{eqnarray}
Then, for given  $N$, the time-averaging MSE is written as
\begin{eqnarray}\label{MSE_s_y_t}
\overline{\rm MSE} =  \frac{1}{N}\sum_{n\in\cal N} {\rm MSE}[n].
\end{eqnarray}
\subsection{Problem Formulation}
In this paper, we aim to minimize $\overline{\rm MSE}$ in \eqref{MSE_s_y_t}, by jointly optimizing the transmit power  $\{ p_{k}[n],\forall n\in\mathcal{N}, \forall k\in\mathcal{K}\}$ at sensors, the denoising factors $\{\eta[n],\forall n\in\mathcal{N}\}$  at the UAV, and UAV trajectory $\{\bm q[n],\forall n\in\mathcal{N}\}$.
The optimization problem  is formulated as 
\begin{subequations}
\begin{eqnarray}
\mathscr{P}:\mathop{\text{minimize}}_{\substack{\{ p_k[n]\},\{\eta[n]\},\\ \{\bm q[n]\}} } &&\overline{\rm MSE}\nonumber \\
\text{subject to}
 &&0\leq p_k[n]\leq P_k,  \forall k,  \forall n, \label{p1001} \\
 &&0\leq\frac{1}{N}\sum_{n=1}^N p_k[n]\leq \bar{P}_k,  \forall k, \label{p1002}\\
 &&\eta[n] \geq 0, \forall n,\label{p1003} \\
&& \|\bm q [n]-\bm q[n-1]\|_2 \leq V_{\max}\delta, \nonumber\\
&&\ \ \ \ n = 1, \ldots, N,\label{p1004}\\
&& \bm{q}[0] = [x_0, y_0],  \label{p1005}\\
 &&\bm{q}[N] = [x_0, y_0].  \label{p1006} 
\end{eqnarray}
\end{subequations}
Although all the constraints in Problem $\mathscr{P}$ are convex, it is challenging to solve  $\mathscr{P}$ due to the non-convex objective function  since the transmit power  $\{p_{k}[n]\}$, denoising factors $\{\eta[n]\}$, and UAV trajectory $\{\bm q[n]\}$ are highly coupled across different time slots. 
In general, there is no standard method for solving such non-convex optimization problems optimally. 
To address these challenges, in the next section, we apply the block coordinate descent (BCD) \cite{xu2013block} and successive convex  approximation (SCA) \cite{marks1978general} techniques to solve problem $\mathscr{P}$.

\section{Proposed Algorithm}
In this section, we propose an efficient iterative algorithm for  problem $\mathscr{P}$ by applying  the BCD\cite{xu2013block} and SCA \cite{marks1978general} techniques. 
Specifically, we optimize one of variables $\{\bm q[n]\}$, $\{p_k[n]\}$, and $\{\bm \eta[n]\}$ and fix others in an alternating manner.
\subsection{Denoising Factor Optimization}
In this subsection, we reformulate problem $	\mathscr{P}$ by optimizing  $\{ \eta[n]\}$  under given $\{\bm p_k[n]\}$ and $\{\bm q[n]\}$  as
	\begin{eqnarray}
	\mathscr{P}_{1.1}:\mathop{\text{minimize}}_{\{\eta[n]\geq 0\}} &&\sum_{n\in\cal N}\!\Bigg(\!\sum \limits_{k\in{\cal K}}\Bigg(\frac{\sqrt{p_{k}[n]}\big|h_{k}[n]\big|}{\sqrt{\eta[n]}}\!-\!1\!\Bigg)^2\!+\!\frac{\sigma^2}{\eta[n]}\!\Bigg). \nonumber
	\end{eqnarray}
Problem $\mathscr{P}_{1.1}$  can be decoupled into $N$ subproblems each for optimization $\eta[n]$ to minimize the MSE at one time slot. The $n$-th subproblem is written as
\begin{eqnarray}\label{n-th eta}
\mathop{\text{minimize}}_{\eta[n]\geq 0} &&\sum \limits_{k\in{\cal K}}\Bigg(\frac{\sqrt{p_{k}[n]}\big|h_{k}[n]\big|}{\sqrt{\eta[n]}}\!-\!1\!\Bigg)^2\!+\!\frac{\sigma^2}{\eta[n]}\!.
\end{eqnarray}
Let $\nu[n] = 1/\sqrt{\eta[n]}$,  then problem $\eqref{n-th eta}$ can be transformed to a convex quadratic problem as
\begin{eqnarray}\label{n-th nu}
\mathop{\text{minimize}}_{\nu[n]\geq 0}  \sum \limits_{k\in{\cal K}}\left(\sqrt{p_{k}[n]}\big|h_{k}[n]\big|\nu[n]-1\right)^2+\sigma^2(\nu[n])^2.
\end{eqnarray}

By setting the first derivative of the objective function in problem $\eqref{n-th nu}$ to be zero, we can obtain the optimal solution
	\begin{eqnarray} \label{solution denoise}
	\eta^{\star}[n]  =  \Bigg( \frac{\sigma^2+ \sum_{k\in{\cal K}} {p}_{k}[n]\big|h_{k}[n]\big|^2}{\sum_{k\in{\cal K}}\sqrt{{p}_{k}[n]}\big|h_{k}[n]\big|}\Bigg)^2.
	\end{eqnarray}

\subsection{Transmit Power  Optimization}
In this subsection, we present the solution to problem $	\mathscr{P}$ by optimizing  $\{ p_k[n]\}$  under given  $\{\bm q[n]\}$ and $\{\eta[n]\}$, which is formulated as
\begin{eqnarray}
	\mathscr{P}_{1.2}:\mathop{\text{minimize}}_{\{ p_k[n]\}\}} &&\sum_{n\in\cal N}\sum \limits_{k\in{\cal K}}\Bigg(\frac{\sqrt{p_{k}[n]}\big|h_{k}[n]\big|}{\sqrt{\eta[n]}}\!-\!1\!\Bigg)^2\nonumber \\
	\text{subject to} 
&&\text{constraints} \ \eqref{p1001}, \eqref{p1002}, \nonumber
\end{eqnarray}
where the constant term $\sigma^2/\eta[n]$  is ignored in the objective function.
In this case, we decompose problem $\mathscr{P}_{1.2}$ into following $K$ subproblems  for optimizing $p_k[n]$, $\forall n \in \mathcal{N}$   to minimize the MSE at one sensor,
	\begin{eqnarray}\label{k-th p}
	\mathop{\text{minimize}}_{\{ p_k[n], \forall n\in \mathcal{N}\}} &&\sum_{n\in\cal N}\Bigg(\frac{\sqrt{p_{k}[n]}\big|h_{k}[n]\big|}{\sqrt{\eta[n]}}-1\Bigg)^2 \nonumber \\
\!\!\!\text{subject to}&&\text{constraints} \ \eqref{p1001}, \eqref{p1002}.
	\end{eqnarray}
Problem \eqref{k-th p} is a convex linearly constrained quadratic program (QP) that can be efficiently solved by standard convex optimization solvers such as CVX \cite{Grant2014CVX}.
\begin{algorithm}[t]
	\caption{Proposed  Algorithm for Solving Problem  $\mathscr{P}$}
	\begin{algorithmic}[1]
		\STATE {\textbf{Input}}:  $T$, $K$, $\{P_k\}$, $\{\bar{P}_k\}$, accuracy  $\epsilon$
		\STATE {{Initialize}}:  initial trajectory $ \{\bm q^0[n]\}$ \\ and initial transmit power $\{p_k^0[n]\}$
		\STATE Let $r = 0$. Set $R^0 = 1$.
		\REPEAT
		\STATE Set $r = r+1$.
		\STATE	Given  $ \{\bm q^{r-1}[n]\}$ and $\{p_k^{r-1}[n]\}$, solve  $\mathscr{P}_{1.1}$ \\ to  update   $\{\eta^{r}[n]\}$  based on  expressions \eqref{solution denoise}.
		\STATE Given $ \{\bm q^{r-1}[n]\}$ and $\{\eta^{r}[n]\}$, solve  $\mathscr{P}_{1.2}$ \\ to update $\{p_k^{r}[n]\}$  based on problems \eqref{k-th p}.
		\STATE Given  $\{p_k^{r}[n]\}$ and $\{\eta^{r}[n]\}$, solve  $\mathscr{P}_{1.3}$ \\to update $\{{ \bm q}^{r}[n]\}$ based on problem \eqref{Trajectory problem3}.
		\STATE Calculate $R^r = \overline{\rm {MSE}}^r$.
		\UNTIL $\frac{R^{r-1} - R^r}{R^r} < \epsilon$.
		\STATE {\textbf{Output}}:  $\{p_k^r[n]\}$, $\{\eta^r[n]\}$, and $\{\bm q^r[n]\}$. 
	\end{algorithmic}
	\label{algo1}
\end{algorithm}  
\subsection{UAV Trajectory Optimization}
In the following, we  optimize the UAV trajectory over $\{\bm{q}[n]\}$ for given $\{ p_k[n]\}$ and $\{\eta[n]\}$. 
We denote
\begin{eqnarray}
f_k[n] &&= \frac{p_{k}[n]\beta_0/\eta[n]}{(H^2+\|\bm{q}[n]-\mathbf{w}_k\|^2)^{\frac{\alpha}{2}}},\\
g_k[n] &&= \frac{2\sqrt{p_{k}[n]}\sqrt{\beta_0}/\sqrt{\eta[n]}}{(H^2+\|\bm{q}[n]-\mathbf{w}_k\|_2^2)^{\frac{\alpha}{4}}}.	
\end{eqnarray}
As a result, the UAV trajectory design is formulated as
\begin{eqnarray}\label{Trajectory problem}	
\mathscr{P}_{1.3}:\mathop{\text{minimize}}_{\{ \bm{q}[n]\}} &&\sum_{n\in\cal N} \sum \limits_{k\in{\cal K}}\Big(f_k[n] - g_k[n]\Big)\nonumber\\
	\text{subject to}
&&\text{constraints} \ \eqref{p1004}, \eqref{p1005}, \eqref{p1006},\nonumber
\end{eqnarray}
where the constant terms $\sigma^2/\eta[n]$ and 1  are ignored.
Since in objective function of Problem $\mathscr{P}_{1.2}$, the terms $f_k[n]$ and $g_k[n]$ are neither convex nor concave with respect to (w.r.t) $\bm{q}[n]$,  problem $\mathscr{P}_{1.2}$ is a non-convex problem. 
In general, there is no efficient method to obtain the optimal solution.
To tackle the non-convexity, in the following, we apply the SCA technique in the trajectory optimization. 
Note that  $f_k[n]$ is convex w.r.t $\|\bm{q}[n] - \mathbf{w}_k\|_2^2$. 
By introducing slack variables $\mathbf{S}=\{s_{k}[n]=\|\bm{q}[n]-\mathbf{w}_k\|_2^2,\forall k,\forall n\} $, problem $\mathscr{P}_{1.3}$ can be reformulated as
\begin{subequations}\label{Trajectory problem2}
	\begin{eqnarray}
	\mathop{\text{minimize}}_{\{ \bm{q}[n]\},\{ s_k[n]\}} &&\sum_{n\in\cal N} \!\sum \limits_{k\in{\cal K}}\!\left(\!\frac{p_{k}[n]\beta_0/\eta[n]}{(H^2+s_{k}[n])^{\frac{\alpha}{2}}}-g_k[n]+1\!\right)\nonumber\\
	\text{subject to}
	&&\text{constraints} \ \eqref{p1004}, \eqref{p1005}, \eqref{p1006},\nonumber\\
	&&s_{k}[n] \leq \|\bm{q}[n]-\mathbf{w}_k\|_2^2,  \forall\, k,\forall\, n,\label{slack1}\\
	&&s_{k}[n] \geq 0,  \forall\, k,\forall\, n, \label{slack2}
	\end{eqnarray}
\end{subequations}
Note that for problem \eqref{Trajectory problem2}, it can be easily verified that  all constraints in \eqref{slack1} can be met with equality, since otherwise the objective value of problem \eqref{Trajectory problem2} can be further decreased by
 increasing $\{s_{k}[n]\}$.
Although reformulated, problem \eqref{Trajectory problem2} is highly challenging to be efficiently solved because of the following two aspects.
First,  the term $-g_k[n]$ is neither convex nor concave w.r.t $\bm{q}[n]$.
Second, even though  $\|\bm{q}[n]-\mathbf{w}_k\|_2^2$ is convex w.r.t $\bm{q}[n]$ in constraints \eqref{slack1}, the resulting feasible set is a non-convex set since the super-level set of a convex quadratic function is not convex in general.

To tackle the non-convexity of $-g_k[n]$ and constraints \eqref{slack1}, the SCA method \cite{marks1978general} can be applied to approximate the original function  by a more tractable function. 
The key observation is that  although  $g_k[n]$  is not convex w.r.t $\bm{q}[n]$, it  is convex w.r.t $\|\bm{q}[n]-\mathbf{w}_k\|_2^2$. 
 Recall that any convex function is globally lower-bounded by its first-order Taylor expansion at any point  \cite{Boyd2004convex}.  
Define $\{\bm{q}^r[n], \forall n\}$ as the given trajectory of UAV in the $r$-th iteration.
 Therefore,  with given local point $\{\bm{q}^r[n], \forall n\}$ in the $r$-th iteration, we obtain the following  lower bound $\hat{g}^{\rm lb}_{k}[n]$ for  $g_k[n]$ as 
\begin{eqnarray}
&&g_k[n] \geq g_k^r[n] \nonumber\\
&& \ +\bigtriangledown_{\bm q[n]} g_k[n]\Big|_{\bm q[n] = \bm q^r[n]}  \Big( \|\bm{q}[n] - \mathbf{w}_k\|_2^2 -  \|\bm{q}^r[n] - \mathbf{w}_k\|_2^2\Big) \nonumber \\ 
&& \ \ \ \triangleq \hat{g}^{\rm lb}_{k}[n], \label{eq37}
\end{eqnarray}
where $\bigtriangledown_{\bm q[n]} g_k[n]\Big|_{\bm q[n] = \bm q^r[n]}  = -\frac{\alpha\sqrt{p_{k}[n]}\sqrt{\beta_0}/\sqrt{\eta[n]}}{2(H^2+\|\bm{q}^r[n]-\textbf{w}_k\|_2^2)^{\frac{\alpha+4}{4}}}$. Note that $\hat{g}^{\rm lb}_{k}[n]$ is concave with respect to $\bm{q}[n]$.

In constraints \eqref{slack1}, since $\|\bm{q}[n]-\mathbf{w}_k\|_2^2$ is a convex function w.r.t $\bm{q}[n]$,
we have the following inequality by applying the first-order Taylor expansion at the given point $\bm{q}^r[n]$,
\begin{eqnarray}
\|\bm{q}[n]-\mathbf{w}_k\|_2^2&&\geq   \|\bm{q}^r[n]-\mathbf{w}_k\|_2^2  \nonumber\\
&& \ \ +  2(\bm{q}^r[n] - \mathbf{w}_k)^T( \bm{q}[n] - \mathbf{q}^r[n] ). \label{eq40}
\end{eqnarray}
With any given local point $\{\bm{q}^r[n], \forall n\}$ as well as  the lower bounds  in  \eqref{eq37} and \eqref{eq40},  problem \eqref{Trajectory problem2} is approximated as the following problem 
\begin{subequations}\label{Trajectory problem3}
	\begin{eqnarray}
\!\!\!\!\!\!\!\mathop{\text{minimize}}_{\{ \bm{q}[n]\},\{ s_k[n]\}} &&\sum_{n\in\cal N} \!\sum \limits_{k\in{\cal K}}\!\left(\!\frac{p_{k}[n]\beta_0/\eta[n]}{(H^2+s_{k}[n])^{\frac{\alpha}{2}}}-\hat{g}^{\rm lb}_{k}[n]+1\!\right)\nonumber\\
	\text{subject to}
	&&\text{constraints} \ \eqref{p1004}, \eqref{p1005}, \eqref{p1006}, \eqref{slack2},\nonumber\\
	&&s_{k}[n] \leq  \|\bm{q}^r[n]-\mathbf{w}_k\|_2^2 \nonumber\\  
	&&\ +2(\bm{q}^r[n] \!-\! \mathbf{w}_k)^T( \bm{q}[n] \!-\! \mathbf{q}^r[n] ),  \forall k,\forall n.\label{TP3cons1}
	\end{eqnarray}
\end{subequations}
Since the objective function is jointly convex w.r.t $\{\bm{q}[n]\}$ and $\{s_k[n]\}$, problem \eqref{Trajectory problem3} is a convex  QCQP problem.
Note that the lower bounds adopted in \eqref{TP3cons1} suggest that any feasible solution of problem \eqref{Trajectory problem3} is also feasible for problem \eqref{Trajectory problem2}. 
Furthermore,    $-\hat{g}^{\rm lb}_{k}[n]$ is adopted to be an upper bound as $-g_k[n]$.
As a result, the optimal objective value obtained from the approximate problem \eqref{Trajectory problem3} in general serves as upper bound of that of problem \eqref{Trajectory problem2}.
\subsection{Overall Algorithm and Computational Complexity}
\begin{figure*}[t]
	\centering
	\subfigure[Convergence behavior of Algorithm \ref{algo1}.]{\includegraphics[width=0.5\columnwidth,height=3.5cm]{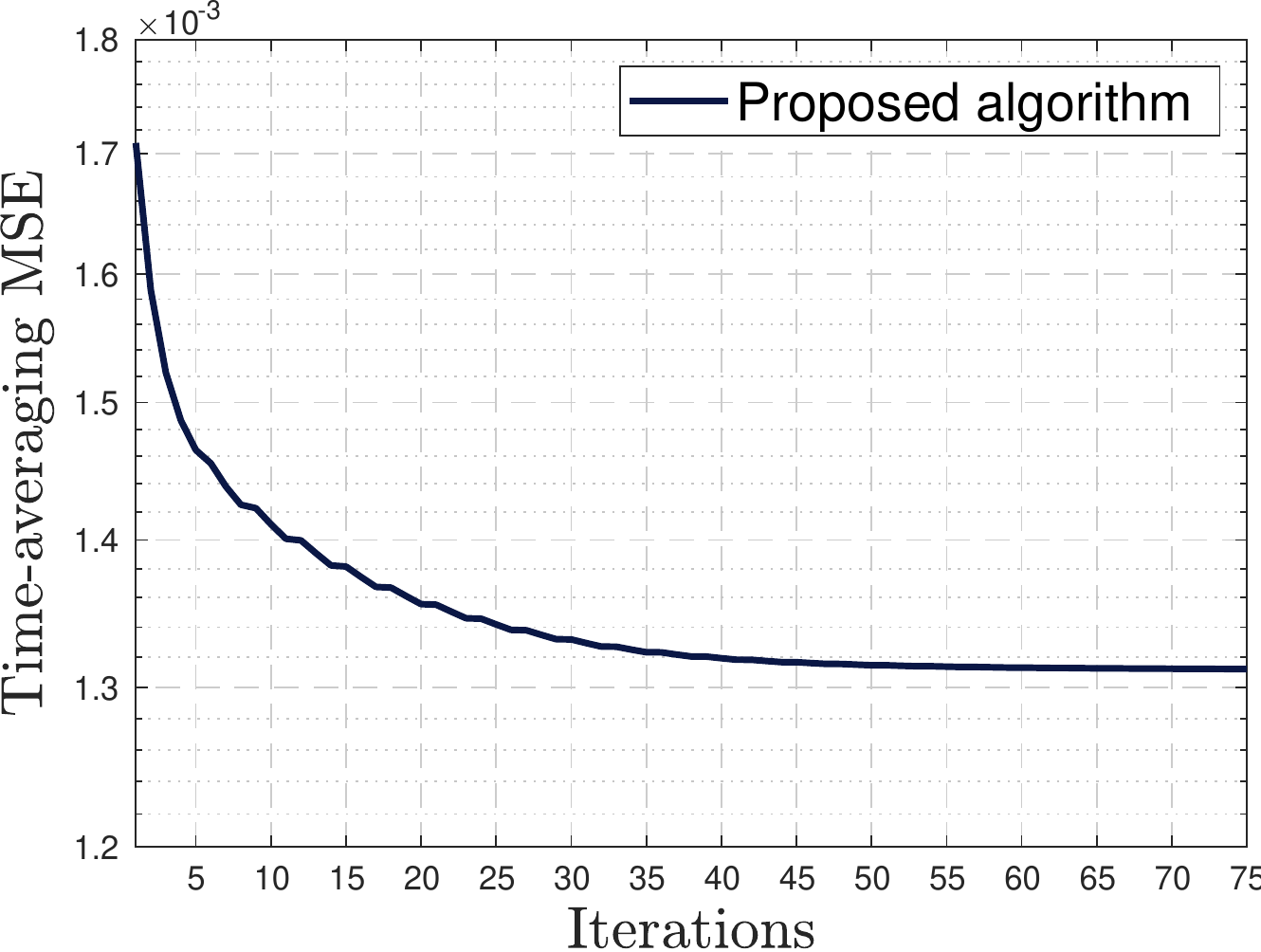}\label{fig:convergence}}
	\subfigure[Optimized UAV trajectories for different time horizons $T$. ]{\includegraphics[width=0.5\columnwidth,height=3.4cm]{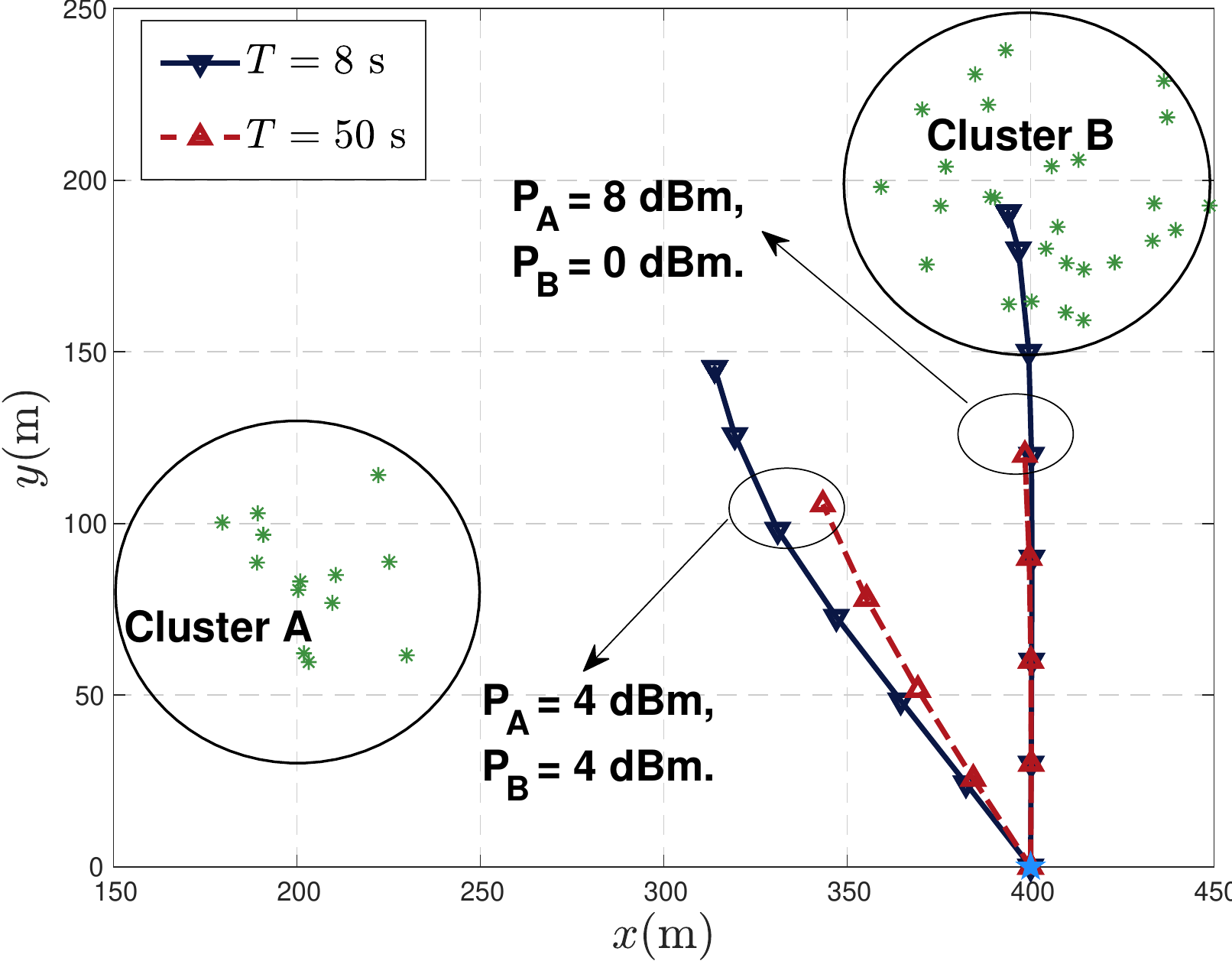}\label{fig:Trajectory}}
	\subfigure[The time-averaging MSE versus mission time.]{\includegraphics[width=0.5\columnwidth,height=3.5cm]{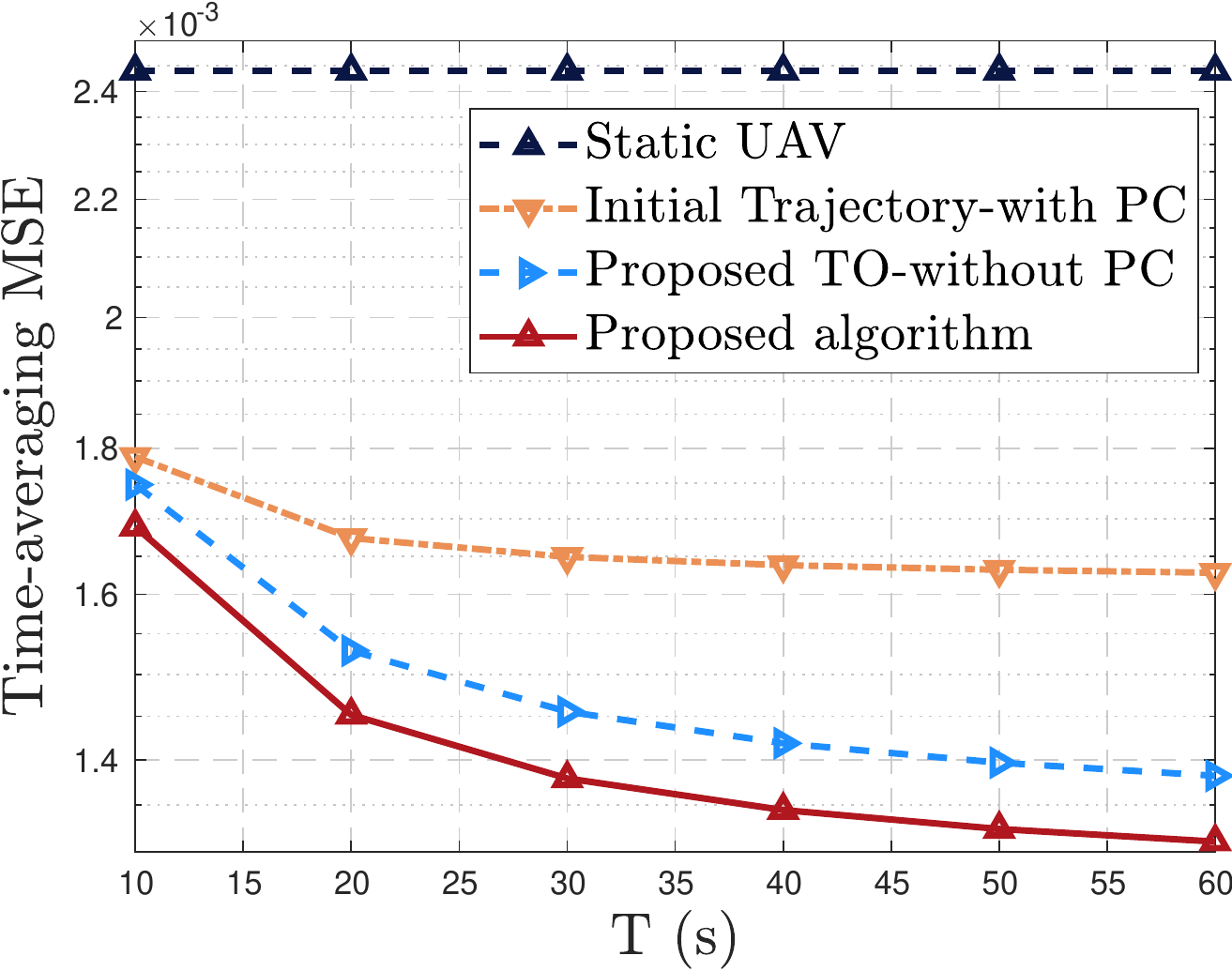}\label{fig:time}}      
	\subfigure[The time-averaging MSE versus noise.]{\includegraphics[width=0.5\columnwidth,height=3.4cm]{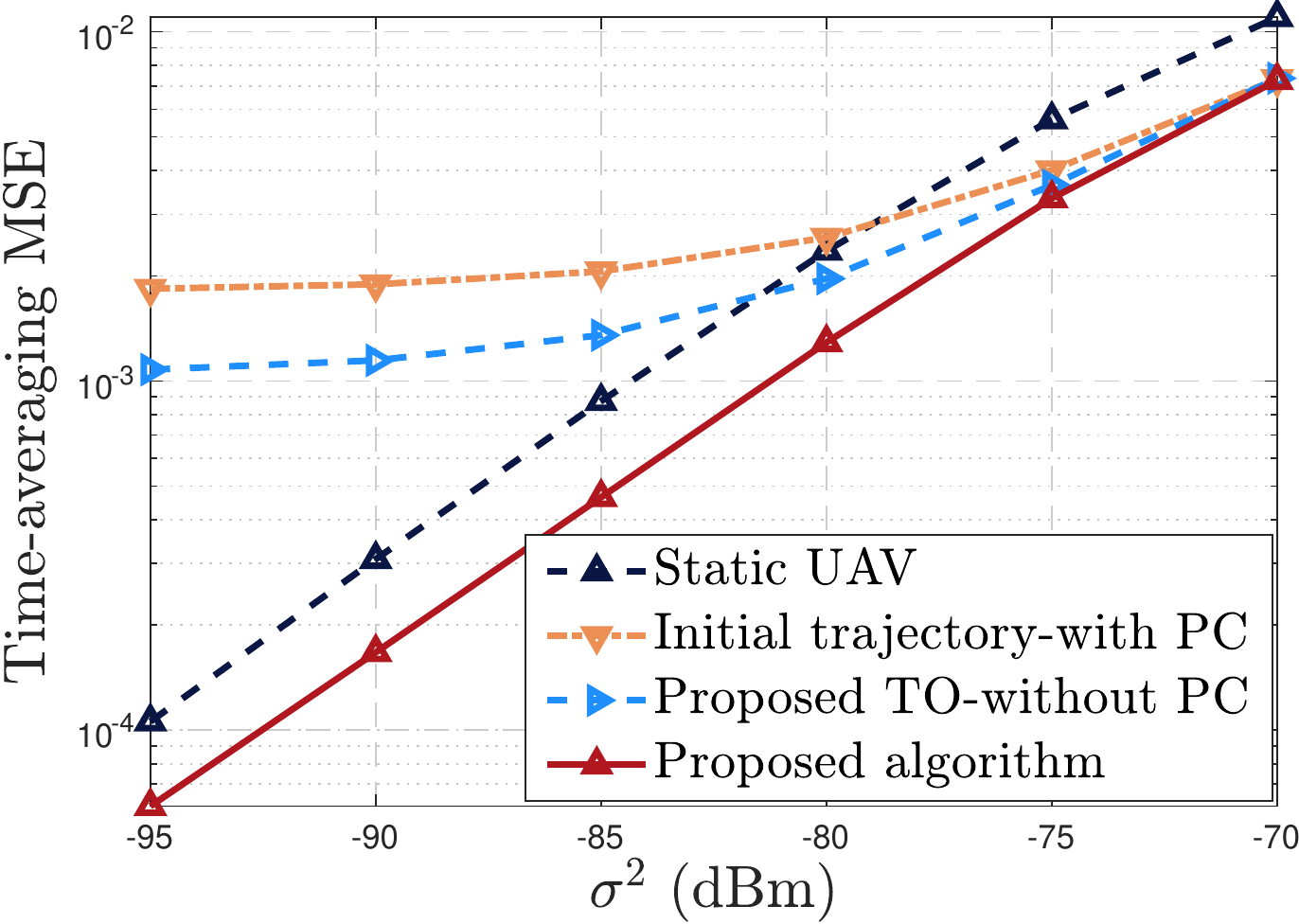}\label{fig:noise}}      
	\vspace{-6mm}
\end{figure*}
In summary, the overall algorithm  adopts the BCD to solve the three sub-problems $\mathscr{P}_{1.1}$, $\mathscr{P}_{1.2}$, and $\mathscr{P}_{1.3}$ alternately in an iterative manner, which can return  a suboptimal solution to problem $\mathscr{P}$.
The details of  the proposed algorithm are summarized in Algorithm \ref{algo1}. 
Problem $\mathscr{P}_{1.1}$ has a closed-form solution at each iteration. Furthermore, in each iteration, problem $\mathscr{P}_{1.2}$ is decoupled into $K$ QP problems \eqref{k-th p} and problem $\mathscr{P}_{1.3}$ is approximated to the QCQP problem \eqref{Trajectory problem3}  by using the SCA technique. 
Since CVX invokes the interior-point method to solve problems \eqref{k-th p} and \eqref{Trajectory problem3}, the total computational complexity of solving problems $\mathscr{P}_{1.2}$ and \eqref{Trajectory problem3} are given by $O(KN^{2.5})$ and $O(K^{2.5}N^{2.5})$, respectively.
Thus, in each iteration,   the computational complexity of the proposed algorithm is given by $O(K^{2.5}N^{2.5})$.
\subsection{Initialization Scheme}
In this subsection, we propose a low-complexity initialization scheme for the trajectory design and power control in Algorithm \ref{algo1}.
Specifically, the initial power of each sensors is set be full power transmission, i.e., $p_k[n] = \bar{P}_k$, $\forall n$.
Furthermore, the UAV first flies straight to the point above the geometric center at its maximum speed, then remains static at that point (if time permits), and finally flies at its maximum speed to reach its final location by the end of the last time slot. Note that if the UAV does not have sufficient time to reach the point above the ground node, it will turn at a certain midway point and then fly to the final location at the maximum speed.
\section{Numerical Results}
In this section, we provide numerical examples to demonstrate the effectiveness of the proposed algorithm. We consider a heterogeneous network,
where $K=40$  sensors are separated into two clusters, i.e., $A$ with 13 sensors and $B$ with 27 sensors. The peak power budget of sensors in the  cluster $A$ and $B$ are denoted as $P_A$ and $P_B$, respectively. 
The sensors in cluster $A$ and cluster $B$ are randomly and uniformly distributed in a circle centered  at (200, 80) meter and (400, 200) meter with  radius $50$ meters, respectively.
The UAV is assumed to fly at a fixed altitude $H = 100$ meters, which complies with the rule  that  all commercial UAV should not fly over 400 feet (122 meters) \cite{FAA2016UAV}. 
The maximum speed of the UAV is set as 30 m/s.
  $\delta$ is set as 0.2 s, which satisfies  $\delta V_{\rm max}\ll H$. 
  The receiver noise power is assumed to be $\sigma^2 = -80$ dBm.
  Channel gain at reference distance is set as $\beta_0=-40$ dB.
  The average power budget is set as $\bar{P}_k = \frac{1}{2}P_k$.   
 In addition, algorithm accuracy $\epsilon$ is set as $ 10^{-4}$.

In Fig. \ref{fig:convergence}, we show the convergence behavior of the proposed algorithm  when $T = 50$ s and $P_k$ = 4 dBm for all $k$.  Fig. \ref{fig:convergence} shows that the average MSE achieved by the proposed algorithm decreases quickly  and  converges in a few iterations, which demonstrates the effectiveness of the proposed algorithm for the joint  UAV  trajectory, power control, and denoising factors design.

In Fig. \ref{fig:Trajectory}, we illustrate the  trajectories obtained by the proposed algorithm under different periods $T$ and  power budgets. 
Each trajectory is sampled every one second and the sampled points are marked with $\bigtriangleup$ by using the same colors as their corresponding trajectories. The user locations are marked by green $*$. The predetermined locations of UAV are  blue  $\star$.
It is observed that when $T = 8$ s,  the UAV  flies close to sensors from the initial location to the end of flight at the maximum speed.
As $T$ increases, the UAV  adaptively enlarges and adjusts its trajectory to  move  closer to the  sensors. 
When time horizon $T$ is sufficiently large, i.e., $T = 50$ s, the UAV first flies at the maximum speed to reach a certain location above the  sensors, then remains stationary at this location as long as possible, and finally
goes back to the initial location in an arc path at the maximum speed.
The main reason for this result is that, in the AirComp setup, all the links' channels are dependent on the UAV's location at the same time slot. 
The closer the UAV flies to one particular  sensor, the farther it is away from  other sensors inevitably. 
As a result, these stationary locations strike an optimal balance  enhancing all the links' channels.
In fact, this is also why the UAV follows an arc path rather than the straight path. 
Furthermore, for the diversified transmit power budgets, we observed that the controllable high-mobility UAV  obtains different trajectories, thereby improving the AirComp performance.

Fig. \ref{fig:time} shows the  time-averaging MSE  versus different flight periods $T$  achieved by the following schemes when $P_A = P_B = 4$ dBm: 1) Proposed algorithm, which is obtained by Algorithm \ref{algo1}; 2) Proposed TO without PC, where transmit power is obtained by  full power transmission; 3) Initial Trajectory with PC, where UAV trajectory is obtained by the  trajectory initialization in Section III-E; 4) Static UAV, where the UAV is placed at the predetermined position $(400, 0, 100)$ meter and remains static. 
First, it is observed that the average MSE achieved by all the schemes but static UAV decreases as $T$ increases by exploiting UAV mobility.
Besides, the proposed algorithm  has the smallest time-averaging MSE compared to other two benchmarks.
Since the proposed algorithm dynamically strikes a balance between transmit power and all the links'  distances by  fully exploiting the trajectory design and power control.
The above results demonstrate the importance and necessity of the  joint design.

Fig. \ref{fig:noise} shows the robustness against noise of different algorithms when $P_A = 4$ dBm, $ P_B = 8$ dBm, and $T = 50$. 
It is observed that the  time-averaging MSE achieved by all schemes increases as the noise power increasing  while the proposed algorithm outperforms other three benchmarks  throughout the whole noise regime.
Interestingly, with low noise power,  we observe that the performance gap between the proposed algorithm and only either power control or trajectory optimization schemes is very large. 
In all, it demonstrates that the proposed algorithm  is robust to noise.
\section{Conclusion}
In this paper, we considered a  UAV-assisted AirComp system, where the  UAV acting as an aerial  BS is dispatched to aggregate the average function of distributed data. 
Therein, due to its controllable high-mobility and high-altitude,  UAV can move sufficiently close to the sensors with LoS channels and  adaptively strike a balance  between communication distance  and sensors' diversified power budgets, which not only  enhances the signal magnitude alignment but also  mitigates the effect of noise.
We studied  the time-averaging MSE minimization problem, which  jointly designs the UAV trajectory, denoising factors at UAV, and transmit power at sensors.
An efficient iterative algorithm was further presented to solve the resulting non-convex optimization problem by applying the BCD and SCA techniques.
Simulation results showed that significant performance gain  and robustness of  the proposed design as compared to other benchmarks as well as the ability of adaptive trajectory construction.
\bibliographystyle{IEEEtran}
\bibliography{ref} 

\end{document}